\documentclass[fleqn,usenatbib]{mnras}

\usepackage[dvipdfmx]{graphicx}

\usepackage{amsmath}
\usepackage{color}

\usepackage[T1]{fontenc}
\usepackage{aecompl}
\usepackage{yfonts}

\newcommand{\lesssim}{\,\raisebox{-0.4ex}{$\stackrel{<}{\scriptstyle\sim}$}\,}
\newcommand{\gtrsim}{\,\raisebox{-0.4ex}{$\stackrel{>}{\scriptstyle\sim}$}\,}

\begin{document}

\title[]{Characterization of the velocity anisotropy of accreted globular clusters}

%  \author[P. Bianchini et al.]{P. Bianchini$^{1,} $\thanks{E-mail:
%bianchip@mcmaster.ca}\thanks{CITA National Fellow} et al.
 \author[P. Bianchini et al.]{P. Bianchini\thanks{E-mail:
bianchip@mcmaster.ca}\thanks{CITA National Fellow},
A. Sills, M. Miholics
\\
Department of Physics and Astronomy, McMaster University, Hamilton, Ontario, L8S 4M1, Canada\\
}

\date{}
\maketitle

\begin{abstract}
Galactic globular clusters (GCs) are believed to have formed in-situ in the Galaxy as well as in dwarf galaxies later accreted onto the Milky Way. However, to date, there is no unambiguous signature to distinguish accreted GCs. Using specifically designed $N$-body simulations of GCs evolving in a variety of time-dependent tidal fields (describing the potential of a dwarf galaxy-Milky Way merger), we analyze the effects imprinted to the internal kinematics of an accreted GC. In particular, we look at the evolution of the velocity anisotropy. Our simulations show that at early phases, the velocity anisotropy is determined by the tidal field of the dwarf galaxy and subsequently the clusters will adapt to the new tidal environment, losing any signature of their original environment in a few relaxation times. At 10 Gyr, GCs exhibit a variety of velocity anisotropy profiles, namely, isotropic velocity distribution in the inner regions and either isotropy or radial/tangential anisotropy in the intermediate and outer regions. Independently of an accreted origin, the velocity anisotropy primarily depends on the strength of the tidal field cumulatively experienced by a cluster. Tangentially anisotropic clusters correspond to systems that have experienced stronger tidal fields and are characterized by higher tidal filling factor, $r_{50}/r_j\gtrsim0.17$, higher mass loss $\gtrsim60\%$ and relaxation times $t_{rel}\lesssim10^9$ Gyr.
Interestingly, we demonstrate that the presence of tidal tails can significantly contaminate the measurements of velocity anisotropy when a cluster is observed in projection.
Our characterization of the velocity anisotropy profiles in different tidal environments provides a theoretical benchmark for the interpretation of the unprecedented amount of three-dimensional kinematic data progressively available for Galactic GCs.

\end{abstract}

%---------------------------------------------------------------------------------------------------------------------------%
%  			(1)	INTRODUCTION
%---------------------------------------------------------------------------------------------------------------------------%
\section{Introduction}

In recent years, evidence has accumulated that the globular cluster (GC) population of the Milky Way (MW) (and other massive galaxies) was assembled from a primordial population of GCs formed in-situ during the early epoch of galaxy formation as well as from a population of GCs accreted during galaxy mergers (e.g. \citealp{MarinFranch2009,Forbes2010,Keller2012}). 

The main evidence of the dichotomy of accreted and in-situ GCs comprises the observations of ongoing stripping of dwarf galaxies, the presence of stellar streams both in the MW and in external galaxies, and the spatial coincidence of outer halo GCs with stellar streams and overdensities (e.g., \citealp{Mackey2004,Mackey2010,Jennings2015}). A primary example is the ongoing stripping of the Sagittarius dwarf galaxy in the MW with GCs being accreted from it (see e.g. the GC M54, \citealp{Ibata1995}, and other associated GCs, such as Ter 8, Arp 2, NGC 5634, Whiting 1 and Ter 7, \citealp{Bellazzini2003,Law2010}). Additional evidence comes from the observations of the bifurcation of the age-metallicity relation of galactic GCs (e.g., \citealp{MarinFranch2009,Forbes2010,Leaman2013}) and from the studies of the distinct properties of metal-poor/blue and metal-rich/red GCs in external galaxies (e.g. \citealp{Pota2013,Harris2017}).

Understanding the exact proportion of GCs that formed in-situ compared to those accreted is fundamental, since it can provide a direct probe of the accretion history of the MW and trace the hierarchal formation of structures in the Universe. Recent studies have been devoted to understand which properties an accretion process could imprint on the dynamical evolution of a GC, in particular related to the change of clusters' morphology due to a changing of galactic potential and the connected tidal field. \citet{Miholics2014} simulated GCs undergoing an instantaneous change of tidal environment, from that of a dwarf galaxy to the one of the MW, and showed that the cluster's size will promptly adapt in a few relaxation times to the new potential. This result was confirmed by the additional simulations of \citet{Miholics2016}, where GCs experience a time-dependent tidal field from an idealized dwarf galaxy-MW merger. 

Moreover, \citet{Bianchini2015b} have focused on the study of clusters that evolved for many relaxation times in the extreme condition given by a compressive tidal field, an environment that can be provided by the central potential of a dwarf galaxy and that keeps the clusters into a super-virialized state. The clusters were later released to isolation, mimicking an accretion of a dwarf galaxy onto the outer halo of the MW. They found that the clusters expand in response to the changing potential but never become as extended as they would if they evolved solely in isolation. Additional studies on the transition from compressive to extensive tidal environments, in presence of stellar evolution, have been recently carried out by \citet{Webb2017}. The authors find that if the clusters experience the compressive tides only during the early evolutionary stages, right after the transition to extensive tides, they can be less massive, more extended and more tidally filling than clusters that experienced solely an extensive tidal field.

These works ultimately suggest that an accretion process is unlikely to imprint long-lasting distinct signatures on the morphology of GCs that could help discriminate accreted from in-situ systems. For this reason, in this paper we wish to undertake a complimentary approach, investigating for the first time the effect of an accretion process on the internal kinematics of clusters that experienced an evolution in a dwarf galaxy-MW merger. 

Our focus on the internal kinematics is motivated by the recent  boost in both quality and quantity of the kinematic observations. In particular, line-of-sight velocity studies provide high-accuracy measurements for both the central and intermediate regions of GCs, thanks to a combination of both integrated-light spectroscopy (e.g. \citealp{Luetzgendorf2013,Fabricius2014}) and resolved spectroscopy (e.g. \citealp{Lanzoni2013,Lardo2015,Kamann2016}). Moreover, a significant improvement was recently possible thanks to \textit{Hubble Space Telescope} (\textit{HST}) proper motions, providing high-precision measurements of the two-dimensional velocities for $\gtrsim10\,000$ stars for selected GCs (e.g., \citealp{Richer2013} and HSTPROMO data sets, \citealp{Bellini2014}), and allowing to directly measure properties like the mass-dependence of the kinematics and the anisotropy in the velocity space (\citealp{Watkins2015,Watkins2015b,Baldwin2016,Bianchini2016a}). In addition to this, the forthcoming data from the ESA Gaia mission will further deliver groundbreaking kinematic measurements especially in the non-crowded regions of GCs (e.g. \citealp{Pancino2017}). 

In order to provide a theoretical benchmark for the interpretation and understanding of this unprecedented amount of kinematic measurements, we exploit the dynamical simulations developed by \citealp{Miholics2016} to specifically study the evolution of the velocity anisotropy of GCs.
The development of velocity anisotropy has been interpreted as the combined result of the formation mechanism of GCs (e.g. violent relaxation, \citealp{Lynden-Bell1967}) and their subsequent long-term dynamical evolution. 
Violent relaxation can imprint a variable flavour and degree of anisotropy, depending on the absence or presence of a tidal field during the phase of dissipationless collapse (see \citealp{vanAlbada1982,Vesperini2014}, respectively). In addition, dynamical evolution can also naturally determine the development of radial anisotropy, especially in the post-core-collapse phase (due to the ejection of stars from the core to the halo preferentially on radial orbits, see \citealp{Spitzer1972,Giersz1994,GierszHeggie1996}) or quench such a growth in case of the presence of a tidal field (e.g. see \citealp{GierszHeggie1997,Takahashi1997}).

More recently, \citet{Tiongco2016} and \citet{Zocchi2016} have studied the evolution of the velocity anisotropy of GCs in a tidal field, confirming that in case of tidally underfilling clusters, the radial anisotropy developed in the earlier stages is slowly suppressed and the systems move toward isotropic velocity distributions. Moreover, \citet{Tiongco2016} showed that clusters initially more underfilling develop stronger radial anisotropy during their long term evolution. However, it has been shown that the tidal field can also have the effect of preferentially stripping stars on radial orbits, making the systems tangentially anisotropic (e.g. \citealp{Takahashi2000,BaumgardtMakino2003}). This indicated that the strength of the tidal field can play a significant role in shaping the particular flavour of velocity anisotropy (\citealp{Sollima2015}).
The goal of our study is to further develop these earlier results, including the cases in which the tidal field is time-dependent and following specifically the evolution of the velocity anisotropy for accreted GCs, in search for a fossil record of accretion.

Our paper is structured as follows. In Section \ref{sec:2} we introduce our set of GCs simulations evolved in a time-dependent tidal field representing a dwarf galaxy-MW merger. In Section \ref{sec:3} we explore the long-term evolution of the velocity anisotropy and its radial dependence. In Section \ref{sec:tidal_strength} we analyze the specific role of the tidal field strength experienced by a cluster in shaping the velocity anisotropy and in Section \ref{sec:5} we present the effects on the outskirts of the clusters. Finally in Section \ref{sec:discussion} we discuss our conclusions.

%---------------------------------------------------------------------------------------------------------------------------%
%  			(2)	Simulations: description and construction profiles
%---------------------------------------------------------------------------------------------------------------------------%
\section{Simulations}
\label{sec:2}

%------------------------------------------------------------------%
\begin{table*}
\tabcolsep=0.20cm
\begin{center}
\caption{Initial and final (10 Gyr) conditions of the set of simulations divided for accreted GCs (Dwarf Evaporates, first 4 rows, and Dwarf Falls scenarios, following 7 rows), GCs evolved only in the MW potential and GCs evolved only in a dwarf galaxy potential. We further report the distance of the dwarf galaxy (and GCs) to the MW centre $d_{MW}$, the distance of the cluster to the centre of the dwarf $d_{DW}$, the mass of the dwarf $M_{DW}$; the half-mass radius $r_{50}$, the Jacobi radius $r_j$ and the filling factor $r_{50}/r_j$ of the clusters.}
\begin{tabular}{lllllllcrrlrrr}
\hline\hline
 &\multicolumn{6}{c}{initial}&\hspace{1cm}&\multicolumn{6}{c}{final}\\
 & $d_{MW}$& $d_{DW}$& $M_{DW}$ & $r_{50}$& $r_j$& $r_{50}/r_j$& &$d_{MW}$& $d_{DW}$& $M_{DW}$& $r_{50}$& $r_j$& $r_{50}/r_j$\\

%&&&&& $[R/R_h]$\\
\hline
Accreted& kpc & kpc & $M_\odot$& pc & pc & & &kpc & kpc & $M_\odot$& pc & pc & \\

\hline
DWL-MW10-evap&	10 & 4 & $10^{10}$ & 3.2 & 34.74 & 0.09  & & 10 &$-$ & $-$& 6.06 &25.81 & 0.23\\
DWS-MW10-evap&	10 & 4 & $10^{9}$ &	3.2 & 46.82 & 0.07 && 10 & $-$& $-$& 6.17 & 26.00&0.24\\
DWL-MW20-evap&	20 & 4 & $10^{10}$&	3.2 & 38.10 & 0.08  && 20 &$-$ & $-$& 8.11 &48.31 &0.17\\
DWS-MW20-evap&	20 & 4 & $10^{9}$ &	3.2 & 63.28 & 0.05&& 20 &$-$ & $-$& 7.94 &50.09 &0.16\\

DWL-MW10-fal&	50 & 4 & $10^{10}$&	3.2 & 39.64 & 0.08 && 10 &4 &$10^{10}$& 3.47 &13.94 &0.25\\
DWS-MW10-fal&	50 & 4 & $10^{9}$ &	3.2 & 80.20 & 0.04&& 10 &4 &$10^{9}$& 6.08 &27.49 &0.23\\
DWL-MW15-fal&	50 & 4 & $10^{10}$ &3.2 & 39.63 & 0.08&& 15 &4&$10^{10}$& 3.63 &17.56&0.21\\
DWS-MW15-fal&	50 & 4 & $10^{9}$ &	3.2 & 80.20 & 0.04&& 15 & 4&$10^{9}$& 7.57 & 38.25&0.20\\
DWS-MW15-R1.6-fal&50& 4 & $10^{9}$ & 1.6& 80.13 & 0.02&& 15 & 4&$10^{9}$& 7.33 & 35.84&0.21\\
DWS-MW15-R4-fal&	 50& 4 & $10^{9}$ &	 4.0& 79.50 & 0.05&& 15 &4&$10^{9}$& 7.27 &38.49&0.19\\
DWS-MW30-fal&	 50& 4 & $10^{9}$& 3.2& 80.20 & 0.04  && 30 &4&$10^{9}$& 7.51 &58.33&0.13\\

\hline
MW only&&&&&\\
\hline
MW10	&	10 & $-$ &$-$ &3.2 & 38.70 & 0.08 &&10 & $-$ & $-$&6.27 & 26.19 & 0.24\\
MW15	&	15 & $-$ & $-$ &3.2 & 52.28 & 0.06&&15 & $-$ & $-$&7.58 & 38.84 & 0.20\\
MW15-R1.6&	15 & $-$ & $-$ &1.6 & 52.40 & 0.03&& 15 & $-$ & $-$& 7.29 & 36.45 & 0.19\\
MW15-R4&	15 & $-$ & $-$ &4.0 & 52.11 & 0.08&& 15 & $-$ & $-$& 7.24 & 39.11 & 0.20\\
MW20	&	20& $-$ & $-$ &3.2& 65.10 & 0.05&&20 & $-$ & $-$&8.41 & 50.21 & 0.17\\
MW30	&	30 & $-$ & $-$ &3.2 & 88.90 & 0.04&& 30 & $-$ & $-$& 8.79 & 71.07 & 0.12\\
\hline
Dwarf only&&&&&\\
\hline
DWL& $-$	& 4 & $10^{10}$ & 3.2& 39.95 & 0.08 && $-$ & 4 & $10^{10}$& 4.87 & 22.40 & 0.22\\
DWS& $-$& 4 & $10^{9}$ & 3.2& 85.96 & 0.04 && $-$ & 4 & $10^{9}$& 8.54 & 67.01 & 0.13 \\
DWS-R1.6&$-$	& 4 & $10^{9}$ &1.6& 86.24 & 0.02&& $-$ & 4 & $10^{9}$& 7.98 & 63.60 & 0.13\\
DWS-R4&$-$ & 4 & $10^{9}$ & 4.0& 85.87 & 0.05&& $-$ & 4 & $10^{9}$& 8.44 & 67.55 & 0.13\\

\hline
\end{tabular}
\label{tab:initial}
\end{center}
\end{table*}
%------------------------------------------------------------------%

The simulations used in this work extend those presented in \citet{Miholics2016} using the $N$-body code Nbody6tt \citep{Renaud2011,Renaud2015}. This code is an extension of the code Nbody6 \citep{Aarseth2003,Nitadori2012}, that provides the flexibility of evolving star clusters in arbitrary time-dependent tidal fields. The time-dependent tidal field is specified by the user with a series of tidal tensors for given time steps, evaluated at the cluster's position within the chosen potential. 

The goal was to simulate the evolution of star clusters in the combined potential due to a dwarf galaxy-Milky Way merger. This process is simulated under two simplifying scenarios ("Dwarf Falls" and "Dwarf Evaporates"), that encompass the relevant physical ingredients in play.
In the "Dwarf Falls" scenario, the GC is in circular orbit around a dwarf galaxy (at 4 kpc from its centre) that progressively falls onto the MW. The initial separation between the two galaxies is 50 kpc, and at the time of 3 Gyr the separation starts to decrease at a rate of 10 kpc/750 Myr until the dwarf galaxy (with the GC) reaches the final distance to the MW centre (of either 10, 15 or 30 kpc) at approximately 6 Gyr. 
In the "Dwarf Evaporates" scenario, the GC is initially on a circular orbit at 4 kpc around a dwarf galaxy that is on a circular orbit around the MW centre at the distance of either 10 or 20 kpc. Starting at 3 Gyr the dwarf mass is exponentially decreased until it reaches zero at 6 Gyr and the GC is left orbiting around the MW centre (see details in \citealp{Miholics2016}).
The two scenarios allow us to study in a simplifying way the main processes playing a key role in the accretion of a GC, that is the increasing strength of the MW tidal field relative to the one of the dwarf galaxy, either because of the decreasing distance from the MW centre or because of the disruption of the dwarf galaxy.

In both scenarios, the dwarf is simulated as a point mass galaxy of either $10^9$ $M_\odot$ or $10^{10}$ $M_\odot$ and the MW is modelled as a point mass bulge, \citet{Miyamoto1975} disc and logarithmic halo (see  \citealp{Miholics2014} for details).
All GCs are modelled using N=50,000 stars arranged according to a \citet{Plummer1911} profile. The masses are drawn from a \citet{Kroupa2001} mass function with lower and upper mass limits of 0.1 and 50 $M_{\odot}$ respectively. The masses of the stars are evolved according to the stellar evolution prescriptions implemented in the code. No primordial binaries are considered.
The initial half-mass radius of most of the clusters is 3.2 pc, but we explore different cluster densities by considering two clusters with different initial half-mass radius (4 pc and 1.6 pc for the simulation Dwarf Falls at 15 kpc of the MW centre, see Table \ref{tab:initial}).

In addition to these simulations, we use a comparison set of simulations (\citealp{Miholics2016}) comprising of GCs with the same initial conditions but evolving on a circular orbit solely around the MW potential (at 10, 15, 20 and 30 kpc) and on a circular orbit solely around the dwarf galaxy potential (at 4 kpc, for both the $10^9$ and $10^{10}$ $M_\odot$ dwarfs). These simulations were run using the code Nbody6 \citep{Aarseth1999,Aarseth2003}.

The initial conditions of the simulations are summarized in Table \ref{tab:initial}. Table \ref{tab:initial} further reports the half-mass radius $r_{50}$, Jacobi radius $r_j$ and filling factor $r_{50}/r_j$, both at the initial time and at 10 Gyr. The Jacobi radius identifies the boundary of the cluster and can be considered as the approximate radius outside of which stars feel a stronger gravitational force of the galaxy than the one of the cluster. We compute the Jacobi radii $r_j$ directly from the diagonalized tidal tensor using equations 10 and 17 of \citet{Renaud2011}.
The final Jacobi radius at 10 Gyr is calculated using the average values from 9.9 to 10.1 Gyr, for the non-accreted simulations. In the case of the accreted simulations, given that the clusters are in orbit around a dwarf galaxy and the tidal field changes significantly within a period, we calculate the Jacobi radius as an average over the period (i.e. 237 Myr for a GC in the $10^{10}$ $M_\odot$ dwarf and 748 Myr for the $10^9$ $M_\odot$ dwarf). We notice that the formalism underlying the calculation of the Jacobi radius is strictly valid for the case of circular orbits (\citealp{Renaud2011}). However, we can consider it a good approximation for the initial and final conditions of our clusters, given that at these stages only one galactic potential is dominant (either the one of the dwarf galaxy or the one of the MW) and the orbits of the clusters are effectively circular.

Table \ref{tab:initial} shows that all of our clusters are initially tidally underfilling ($r_{50}/r_j<0.1$) and by 10 Gyr of evolution the majority of them reach a tidally filling configuration (approximately defined as $r_{50}/r_j>0.145$, \citealp{Henon1961,Alexander2013,Zocchi2016}).

The simulations used in the paper include a radial cut-off, removing stars beyond a given radius. The code Nbody6 used for the comparison simulations and Nbody6tt used for the accreted simulations are based on a different prescription for the radial cut-off. 
In the the former, all stars beyond $2\times r_{cut-off}$, with $r_{cut-off}=r_j$ are removed, and in the latter with $r_{cut-off}=10\times r_{50}$. In the following, unless otherwise specified, we exclude from our kinematic analysis all the stars outside the 99\% Lagrangian radius, in order to minimize the differences arising from the different radial cut-off prescriptions.

%------------------------------------------------------------------%
\begin{table}
\tabcolsep=0.3cm
\begin{center}
\caption{Half-mass relaxation time $t_{rel}$ at the initial conditions of the simulations, at the end of the accretion process at 6 Gyr, and at 10 Gyr.}
\begin{tabular}{lccc}
\hline\hline
 &\multicolumn{3}{c}{$t_{rel}$ [Gyr]}\\
 &0 Gyr& 6 Gyr & 10 Gyr\\

\hline
Accreted& & & \\

\hline
DWL-MW10-evap&0.31&0.74&0.69\\
DWS-MW10-evap&0.31	&0.88&0.72\\
DWL-MW20-evap&0.32 &1.11&1.36\\
DWS-MW20-evap&0.31	&1.29&1.41\\

DWL-MW10-fal&0.31	&0.45&0.14\\
DWS-MW10-fal&0.31	&1.23&0.79\\
DWL-MW15-fal&0.32	&0.49&0.17\\
DWS-MW15-fal&0.31	&1.26&1.25\\
DWS-MW15-R1.6-fal& 0.11&1.06&1.08\\
DWS-MW15-R4-fal&0.44 &1.49&1.20\\
DWS-MW30-fal&0.31	&1.36&1.38 \\

\hline
MW only&&&\\
\hline
MW10	&0.31&0.88&	0.75\\
MW15	&0.31&1.14&	1.22\\
MW15-R1.6&0.11	&1.04&1.03\\
MW15-R4&0.44&1.36&1.15\\
MW20	&0.31&1.31&1.54\\
MW30	&0.31&1.44&1.77\\
\hline
Dwarf only&&&\\
\hline
DWL	&0.32 &0.62&0.35\\
DWS  &0.32&1.34&1.60\\
DWS-R1.6& 0.11&1.08&1.32\\
DWS-R4& 0.44&1.64&1.61\\
\hline
\end{tabular}
\label{tab:relax}
\end{center}
\end{table}
%------------------------------------------------------------------%

%---------------------------------------------------------------------------------------------------------------------------%
%  			(3)	time evolution of velocity anisotropy
%---------------------------------------------------------------------------------------------------------------------------%
\section{Time evolution of the velocity anisotropy}
\label{sec:3}

 %---------------------------------------------------------------------------%
\begin{figure*}
\centering
\includegraphics[width=0.95\textwidth]{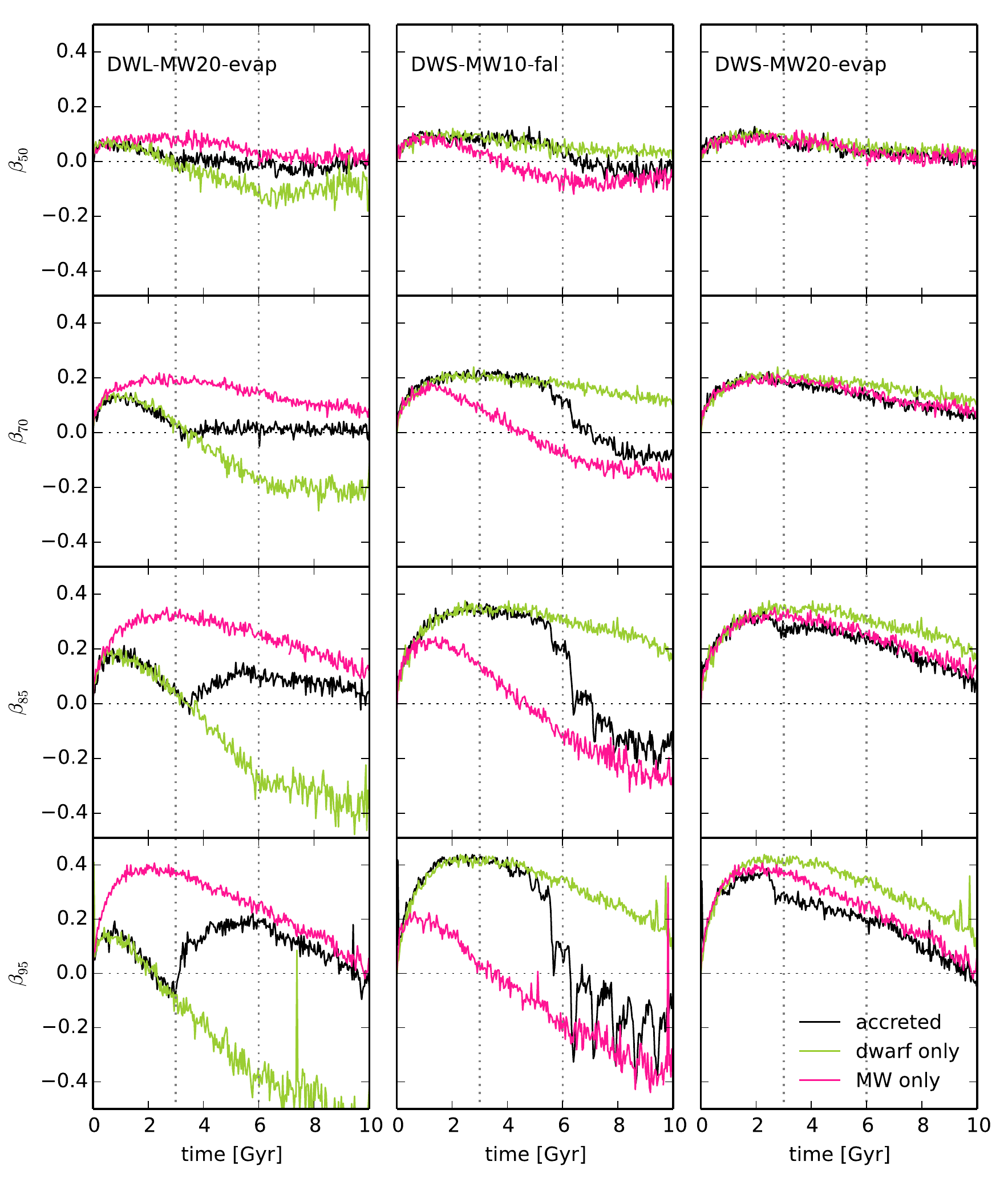}
\caption{Time evolution of the velocity anisotropy parameter $\beta$ calculated at different Lagrangian radii, namely between 40\% and 60\%, 60\% and 80\%, 80\% and 90\% and 90\% and 99\%, denoted as $\beta_{50}$, $\beta_{70}$, $\beta_{85}$, $\beta_{95}$, respectively. The three columns show three different simulations of accreted GCs (black lines) and additionally the corresponding simulations of a GC evolved entirely in the MW (magenta lines) and evolved entirely in a dwarf galaxy (green lines). The vertical dotted lines show the beginning (3 Gyr) and the end of the accretion process (6 Gyr). The strongest differences between the velocity anisotropy of accreted and non-accreted GCs are observable in the outer Lagrangian radii and smooth out in a few relaxation times ($\sim$few Gyr) adapting to the new dominating tidal field. Note that GCs characterized by radial ($\beta>0$), tangential ($\beta<0$) or isotropic ($\beta\simeq0$) velocity distributions are produced.}
\label{fig:timeev}
\end{figure*}
%---------------------------------------------------------------------------% 
 %---------------------------------------------------------------------------%
\begin{figure*}
\centering
\includegraphics[width=1\textwidth]{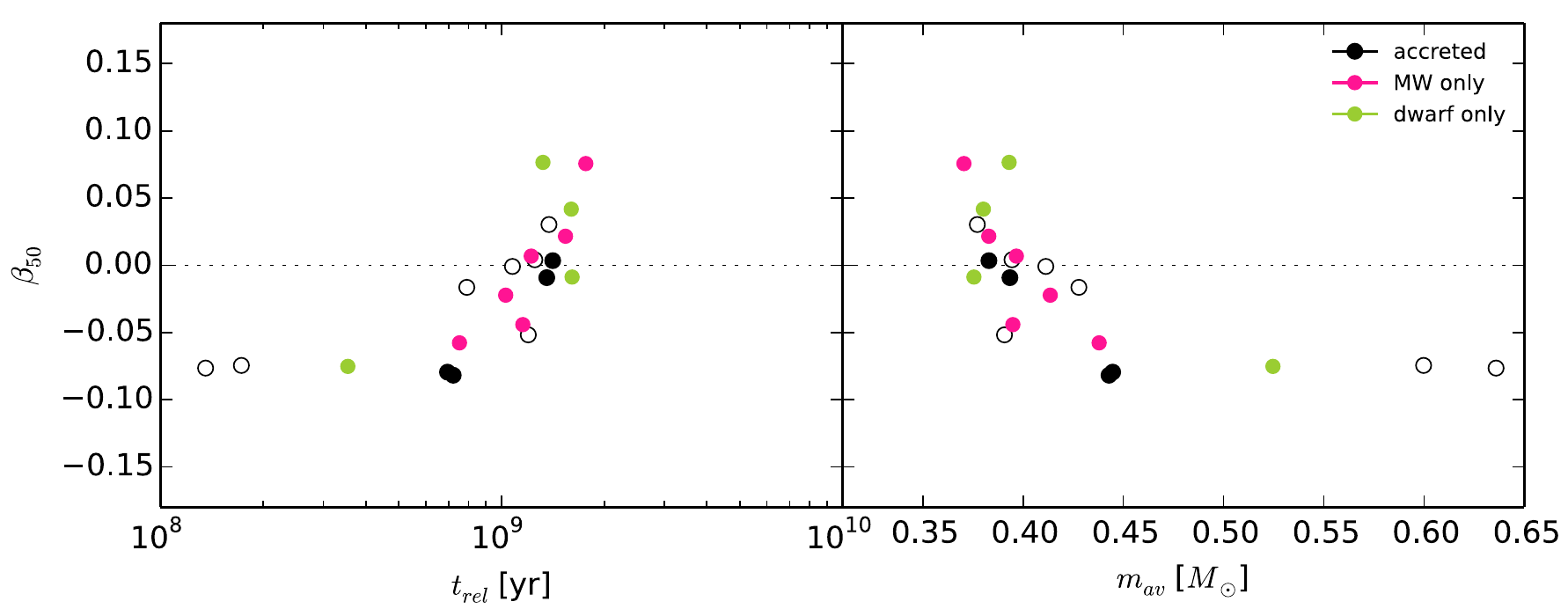}
\caption{\textit{Left panel:} Velocity anisotropy at the half-mass radius $\beta_{50}$ for all the simulations at 10 Gyr as a function of their current half-mass relaxation time $t_{rel}$. Accreted GCs of the type "Dwarf Evaporates" are represented with black dots, accreted clusters of type "Dwarf Falls" are empty circles, GCs evolved solely in the MW are magenta dots and GCs evolved solely in a dwarf galaxy are green dots. Clusters that are dynamically younger (longer relaxation times) are characterized by stronger radial anisotropy; isotropy is reached by clusters with $t_{rel}\simeq10^9$ yr, consistent with the observational finding by \citet{Watkins2015}. \textit{Right panel:} Velocity anisotropy at the half-mass radius at 10 Gyr as a function of the average stellar mass of the clusters. Radial anisotropic systems (i.e., dynamically younger clusters) have lower average stellar masses than systems characterized by isotropy or tangential anisotropy (i.e. dynamically older clusters). This indicates that the tidal field preferentially strips low-mass stars in radial orbits.}
\label{fig:relax}
\end{figure*}
%---------------------------------------------------------------------------% 

For every simulation we analyze the time evolution of the anisotropy in the velocity space using the parameter $\beta$ defined as
\begin{equation}
\beta=1-\frac{\sigma^2_{\phi}+\sigma^2_{\theta}}{2\sigma^2_r},
\end{equation}
with $\sigma_r$, $\sigma_\theta$ and $\sigma_\phi$ the radial and tangential velocity dispersions in a spherical coordinate system relative to the centre of the cluster. Values of $\beta>0$ correspond to a configuration of radial anisotropy,  $\beta<0$ to tangential anisotropy and $\beta\simeq0$ to isotropy.
Since we expect the kinematic properties of a cluster to vary locally with the distance to its centre, we calculate $\beta$ for different Lagrangian radii (radii containing a certain percentage of the cluster's total mass). We exclude from the analysis all stars outside the 99\% Lagrangian radius. This allows us to minimize the contamination from tidal tails and stars that are not bound to the cluster. We consider the Lagrangian radii between 40\% and 60\%, 60\% and 80\%, 80\% and 90\%, and 90\% and 99\%, denoted as $\beta_{50}$, $\beta_{70}$, $\beta_{85}$, $\beta_{95}$, respectively. We will specifically study the kinematic effects in the very outskirts, for the stars beyond the 99\% Lagrangian radius in Section \ref{sec:5}.

Figure \ref{fig:timeev} shows three representative cases of accreted GCs and their corresponding simulations evolved solely in the MW and solely in the dwarf potential. All simulations start with isotropic velocity distribution and gradually reach moderate degrees of radial anisotropy around 1-2 Gyr, corresponding to the expansion phase of the clusters. In this phase, the radial anisotropy is stronger for outer radii. The particular value of velocity anisotropy is determined by the tidal environment of the host dwarf galaxy dominating this earlier evolutionary stage: a less massive dwarf (e.g. DWS-MW10-fall and DWS-MW20-evap in Figure \ref{fig:timeev}) is responsible for a weaker tidal field and gives rise to a more radial anisotropic cluster. Interestingly, the maximum value of $\beta$ is reached when the clusters have experienced a mass loss of 35\%-40\%, for all simulations.

At 3 Gyr the accretion process starts (either as a decrease of the dwarf mass for the Dwarf Evaporates case, or as a decrease of the distance to the MW centre for the Dwarf Falls case) and the clusters begin to adapt to the new tidal field. By the end of the accretion process at 6 Gyr, the accreted clusters can either display lower values of $\beta$ (e.g. first column of Fig. \ref{fig:timeev}), higher values of $\beta$ (e.g. second column of Fig. \ref{fig:timeev}) or comparable values to the corresponding MW GCs. This is particular evident for the intermediate-outer radii, $\beta_{70}$ and $\beta_{85}$.

In the phase immediately following the accretion, differences in velocity anisotropy are observable, however the outcome does not provide a distinctive and unique signature: every single GC will display a particular flavour of velocity anisotropy as a result of its complex dynamical evolution, in particular reflecting the differences in the initial and final tidal environment. 
Specifically, clusters that initially evolved in a stronger tidal field than the one they experience in the MW after the accretion, will display less radial anisotropy than the GCs evolved solely in the MW (e.g. first column Fig. \ref{fig:timeev}, simulation DWL-MW20-evap). Vice versa, clusters that were initially in a weaker tidal field than the one at their final position after the accretion will display more radial anisotropy (e.g. second column Fig. \ref{fig:timeev}, simulation DWS-MW10-fal). In the case in which the initial and final tidal environments are comparable, no differences are present (e.g. third column Fig. \ref{fig:timeev}, simulation DWS-MW20-evap).

Following the dynamical evolution of the clusters until 10 Gyr, the differences in velocity anisotropy are erased and the $\beta$ parameter converges to the values of the corresponding GC evolved in the MW only. This is due to internal relaxation processes, since the clusters have experienced a few relaxation times of evolution in the 4 Gyr following the accretion process (ranging from $\simeq3-10$ $t_{rel}$, with $t_{rel}$ the half-mass relaxation time, see Table \ref{tab:relax}).

In the left panel of Figure \ref{fig:relax} we show the values of the velocity anisotropy at the half-mass radius, $\beta_{50}$, for all our simulations at 10 Gyr, as a function of their current half-mass relaxation time $t_{rel}$. Less relaxed clusters (clusters with longer relaxation times) display radial anisotropy, while more relaxed clusters (clusters with shorter relaxation times) display stronger tangential anisotropy. The right panel of Figure \ref{fig:relax} shows $\beta_{50}$ as a function of the average stellar mass in the clusters, $m_{av}$. Clusters with higher tangential anisotropy are characterized by a higher average stellar mass. This indicates that the tidal field has been efficient in stripping low-mass stars. These stars are set preferentially on radial orbits while they move toward the outskirts of the cluster, because of the effects connected to energy equipartition and mass segregation (e.g. \citealp{Vesperini1997,BaumgardtMakino2003}) that are more efficient for clusters with shorter relaxation times (e.g. \citealp{Bianchini2016b,Bianchini2017}). The clusters will therefore develop tangential anisotropy due to the preferential stripping of low-mass stars on radial orbits. 

An isotropic velocity distribution is reached for clusters with relaxation times of $t_{rel}\simeq10^9$ Gyr. This trend is consistent with the recent observational study of \citet{Watkins2015}, that measured directly the anisotropy of a sample of 22 GCs using \textit{HST} proper motions. Note that no difference in velocity anisotropy is observed between accreted GCs and GCs that did not undergo an accretion process.  

\subsection{Velocity anisotropy profiles}

 %---------------------------------------------------------------------------%
\begin{figure}
\centering
\includegraphics[width=0.48\textwidth]{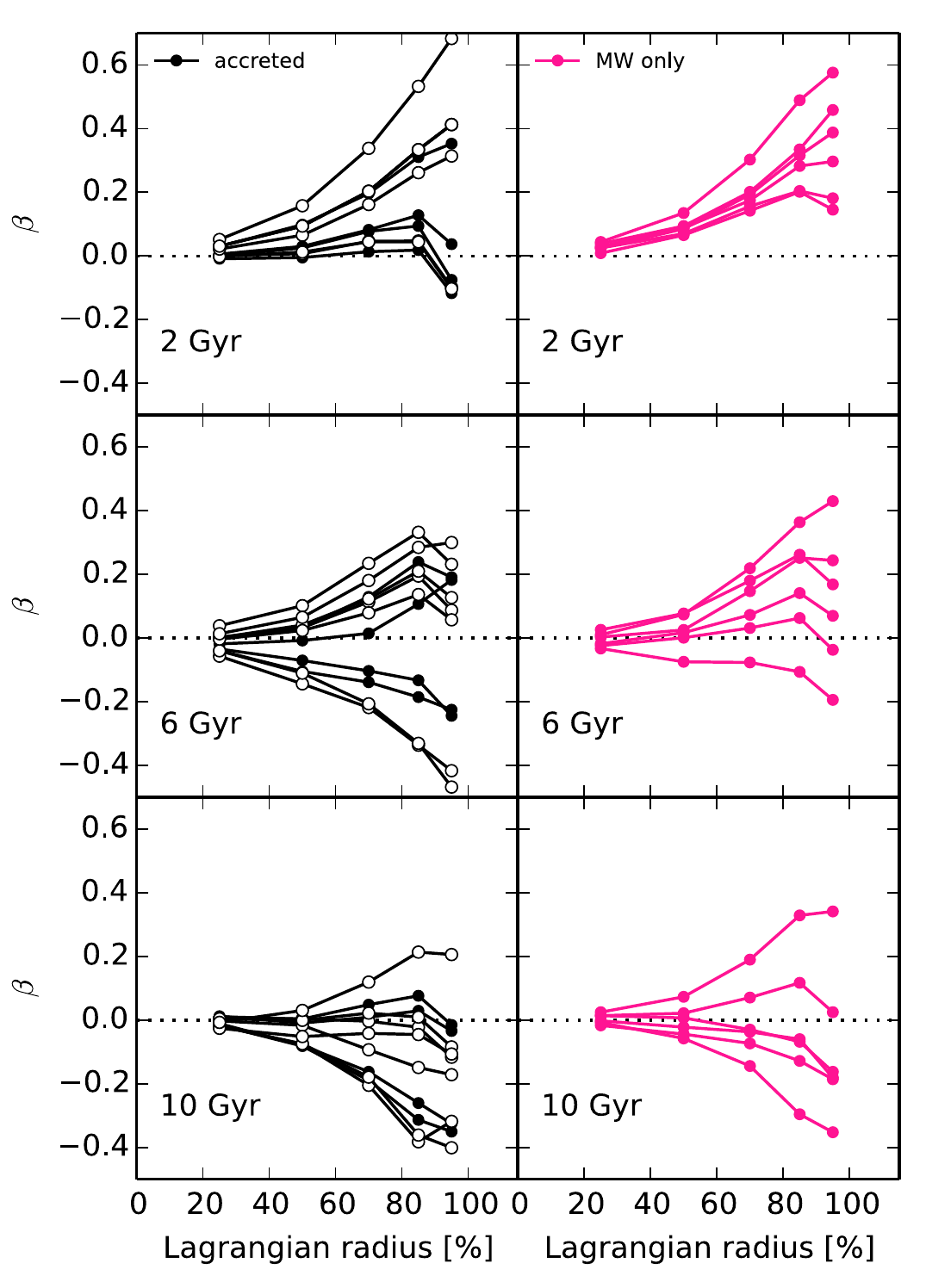}
\caption{Velocity anisotropy profiles of the accreted GC simulations (left panes) and the one evolved solely in the MW (right panels) at 2, 6 and 10 Gyr (corresponding to snapshots before the accretion, right after the accretion and after an evolution of a few relaxation times). The full black circles refer to simulations of type "Dwarf Evaporates", while the empty black circles to the ones of type "Dwarf Falls". Accreted GCs do not show properties of the anisotropy profiles that are distinctive from the ones of GCs evolved solely in the MW .}
\label{fig:profiles}
\end{figure}
%---------------------------------------------------------------------------% 

In Figure \ref{fig:profiles}, we show the velocity anisotropy profiles for all our accreted GCs and the corresponding simulations evolved solely in the MW at three different time snapshots, 2, 6 and 10 Gyr. The values of $\beta$ are computed in Lagrangian radii (see Section \ref{sec:3}), averaging all the snapshots in a time interval of 0.2 Gyr around the selected time snapshot (e.g. between 5.9 Gyr and 6.1 Gyr). The inner most point of each profile correspond to the value of $\beta$ calculated within the 50\% Lagrangian radius. As noted above in Figure \ref{fig:timeev}, earlier snapshots are characterized by radial anisotropy and become progressively more isotropic and tangential at later time steps, while the tidal field becomes more effective in shaping the internal dynamics of the clusters, preferentially removing stars in radial orbits.

This is consistent with recent studies on the evolution of velocity anisotropy profiles by \citet{Zocchi2016} and \citet{Tiongco2016} where the radial anisotropy developed in the earlier phases of evolution is suppressed, especially in the outer parts, due to the effect of the tidal field. Note, however, that the clusters in these previous works only develop isotropic velocity distributions at late time snapshots and do not develop significant tangential anisotropy. Our simulations, instead, give rise to a variety of different profiles, characterized by isotropy in the inner region and either radial anisotropy, tangential anisotropy or isotropy in the intermediate and outer parts, starting from $\simeq r_{50}$. Specifically, the origin of the profiles that become progressively more tangential anisotropic can be understood in term of the strength of the tidal field that the clusters have experienced and by the fact that the clusters reach a tidally filling configuration at 10 Gyr (see Table \ref{tab:initial}). We will show in Section \ref{sec:tidal_strength} that the different values of $\beta$ at 10 Gyr depend on the intensities of the tidal fields that the clusters have cumulatively experienced throughout their evolution (as suggested by \citealp{Sollima2015} evolving two simulations that experienced different tidal field strengths).

Additionally, since the simulations of \citet{Zocchi2016} and \citet{Tiongco2016} do not take into consideration an initial mass function, the stars do not experience mass segregation and energy equipartition. Mass segregation and energy equipartition would make low-mass stars kinematically hotter because of the energy exchanges with high-mass stars. Low-mass stars would therefore reach the boundary of the clusters in radial orbits where they can more efficiently escape. Clusters that experienced mass segregation could therefore exhibit stronger tangential anisotropy. This is consistent with the result from \citet{BaumgardtMakino2003} and the recent result from \citet{Sollima2015}, where tangential anisotropic profiles throughout the cluster's spatial extent are created for a simulation characterized by a initial high filling factor. 

Moreover, we note that the variety of shapes of the velocity anisotropy profiles is recovered even when we analyze the simulations at a fixed relaxation state (number of relaxation times that a clusters has experienced) or at a fixed mass loss, instead of fixing the clusters' age. This supports the idea that the strength of the tidal field that a cluster has experienced is the crucial parameter driving the evolution of the velocity anisotropy profiles.

Finally, Figure \ref{fig:profiles} shows that accreted and MW GCs at 10 Gyr all display similar values of anisotropy and therefore, are not globally distinguishable. We will study the specific shape of the profiles in the outskirts of the clusters (stars beyond the 99\% Lagrangian radius) in Section \ref{sec:5}.

\section{The role of the tidal field}
\label{sec:tidal_strength}

 %---------------------------------------------------------------------------%
\begin{figure*}
\centering
\includegraphics[width=0.98\textwidth]{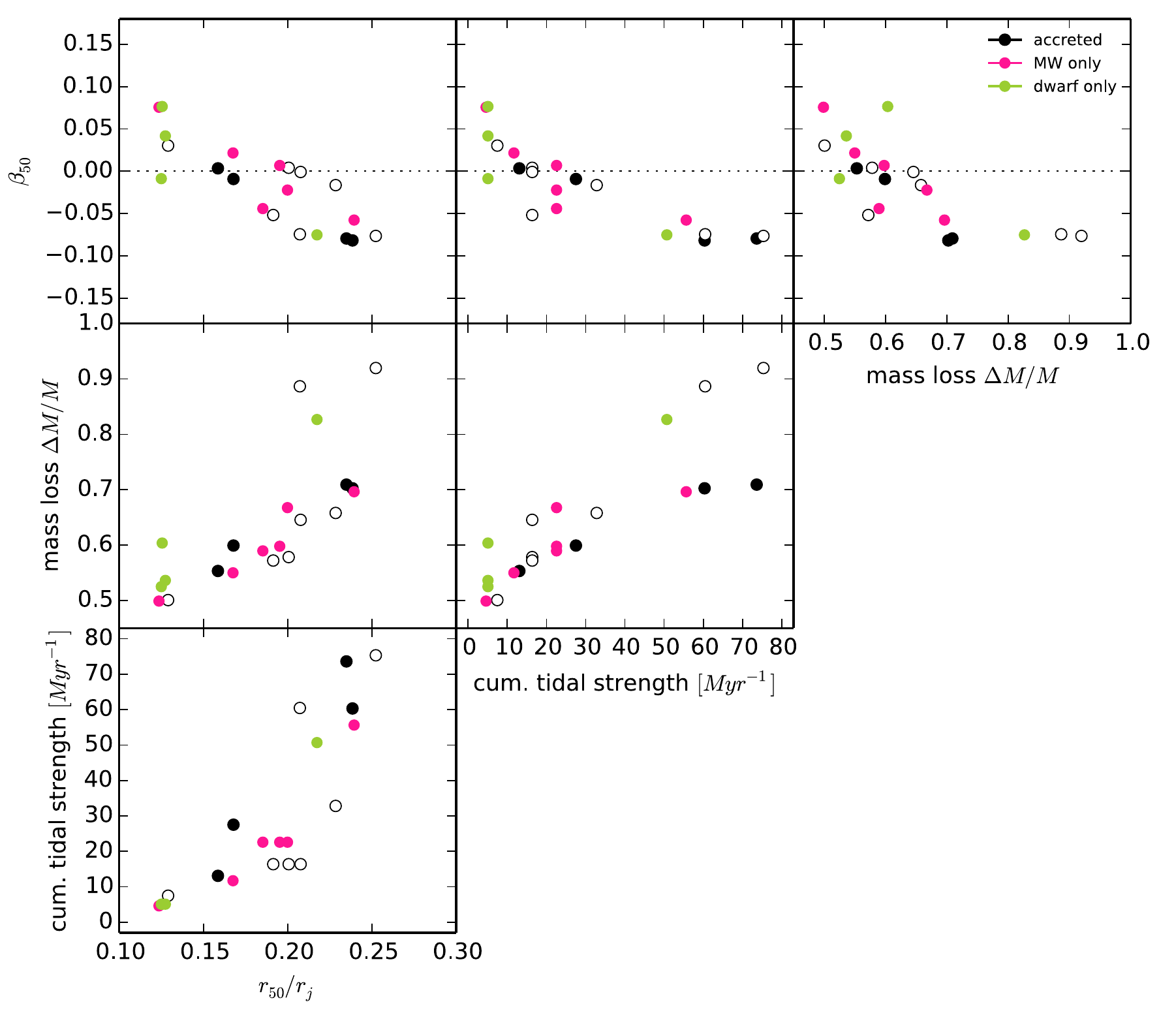}
\caption{Correlations between the velocity anisotropy at the half-mass radius $\beta_{50}$, the filling factor $r_{50}/r_j$, the cumulative strength of the tidal field and the mass loss, calculated at 10 Gyr for our set of GC simulations. The colour coded is as in Figure \ref{fig:relax}, with the full black circles referring to simulations of type "Dwarf Evaporates", while the empty black circles to the ones of type "Dwarf Falls". The cumulative strength of the tidal field is the main factor driving the evolution of the velocity anisotropy: GCs that have experienced a stronger tidal field are characterized by stronger tangential velocity anisotropy and a stronger mass loss. These clusters have all reached a tidally filling configuration characterized by $r_{50}/r_j\gtrsim0.17$ and mass loss $>60\%$. We observe no differences in velocity anisotropy at the half-mass radius between clusters that have been accreted and cluster that evolved solely in the MW.}
\label{fig:correlations} 
\end{figure*}
%---------------------------------------------------------------------------% 

In the previous section, we showed that after 10 Gyr of evolution GCs that underwent an accretion process and GCs that evolved solely in the MW can display a similar wide range of velocity anisotropy, from mildly radial to mildly tangential. In this section we study the main driver of the evolution of the velocity anisotropy parameter $\beta$.

Fixing the age at 10 Gyr, we show in Fig. \ref{fig:correlations} the correlations between the velocity anisotropy at the half-mass radius $\beta_{50}$, the filling factor $r_{50}/r_j$, the cumulative strength of the tidal field and the mass loss experienced by the cluster (defined as the ratio between the mass lost by the cluster and its initial mass, $\Delta M/M$). We define the cumulative strength of the tidal field experienced by a cluster until 10 Gyr using the diagonalized tidal tensors. Given the eigenvalues $\lambda_1\geq\lambda_2\geq\lambda_3$, the value of the leading eigenvalue $\lambda_1$ can be used as a proxy of the tidal field strength at a given time step and is expressed in Myr$^{-2}$ \citep{Renaud2011}. For each simulations, we sum this value for every time step until 10 Gyr and multiply it by the time interval of every time step (the cumulative tidal strength is therefore measured in Myr$^{-1}$). Larger values of this parameter correspond to stronger cumulative tidal field experienced by a cluster. 

Figure \ref{fig:correlations} shows that the velocity anisotropy is strictly connected to the strength of the tidal field experienced by the cluster throughout its life, independently of an accreted origin. Clusters that have experienced stronger tidal fields have also experienced stronger mass loss and are characterized by more tangential velocity distributions due to the preferential stripping of low-mass stars in radial orbits, as noted in Section \ref{sec:3}. The transition between radial and tangential anisotropic velocity distributions takes place at mass losses $\simeq60\%$.
Figure \ref{fig:correlations} also shows the importance of the filling factor in shaping the velocity anisotropy. All clusters with current filling factor $r_{50}/r_j\gtrsim0.17$ are characterized by isotropic/tangential velocity distributions around the half-mass radius.

\section{Velocity anisotropy in the outskirts of the clusters}
\label{sec:5}

 %---------------------------------------------------------------------------%
\begin{figure*}
\centering
\includegraphics[width=0.98\textwidth]{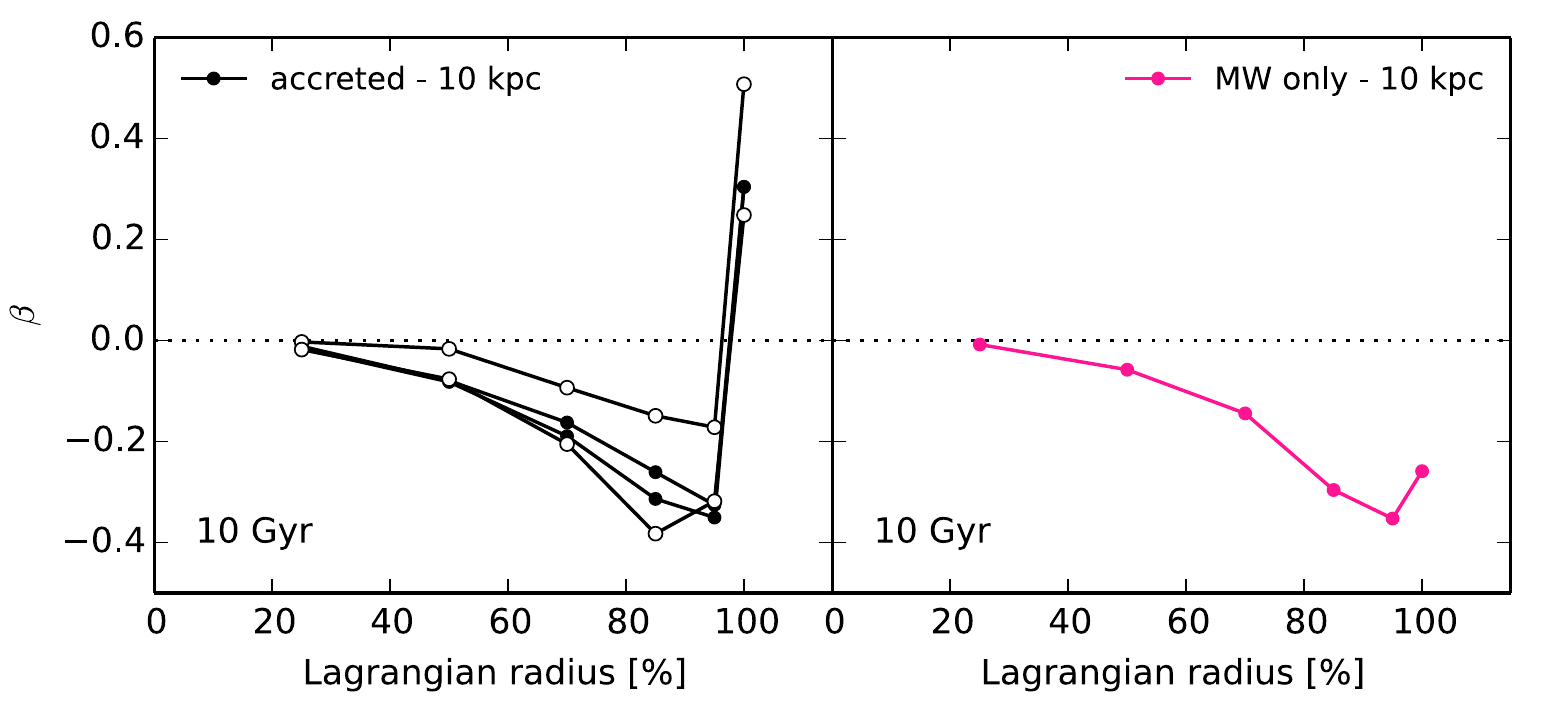}
\caption{Velocity anisotropy profiles for GCs at the distance of 10 kpc at a 10 Gyr snapshot. The full black circles refer to simulations of type "Dwarf Evaporates", while the empty black circles to the ones of type "Dwarf Falls". The anisotropy profiles now include the outer most point corresponding to stars at radii larger than the 99\% Lagrangian radius. The sharp increase of $\beta$ is indicative of the presence of tidal tails.}
\label{fig:10kpc}
\end{figure*}
%---------------------------------------------------------------------------% 

 %---------------------------------------------------------------------------%
\begin{figure*}
\centering
\includegraphics[width=0.9\textwidth]{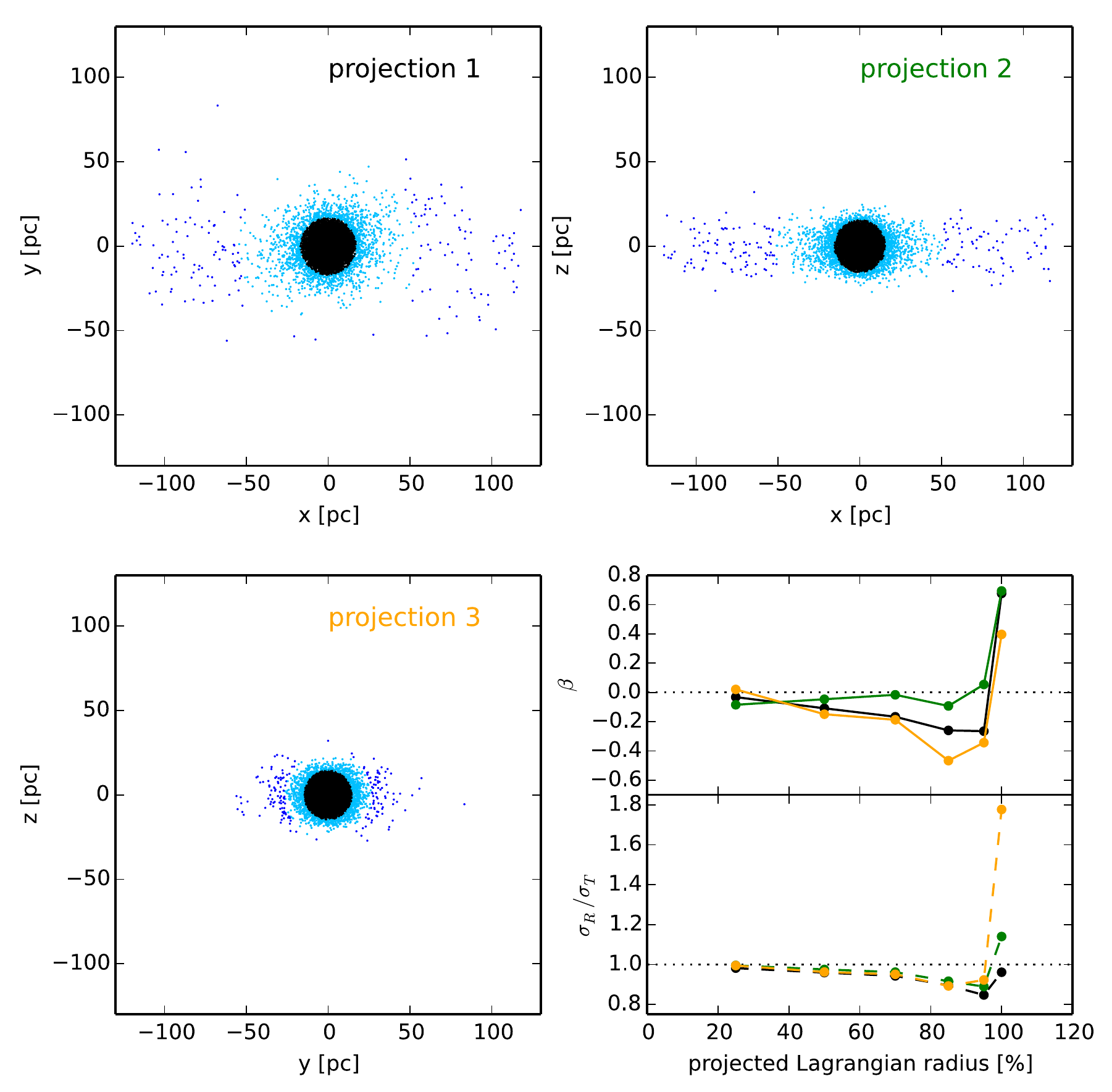}
\caption{Three projected views of the 10 Gyr snapshot of the simulations DWL-MW10-evap. The x-y plane corresponds to the orbital plane of the cluster. The stars are colour coded according to their projected Lagrangian radii: black points are stars within the 90\% projected Lagrangian radius, cyan points are stars between 90 and 99\% projected Lagrangian radii and blue points are stars beyond the 99\% projected Lagrangian radii. This outer radial bin contains most of the stars in the tidal tails. In the bottom right panel we show the velocity anisotropy profiles (both as the $\beta$ parameter and the ratio of the projected radial and tangential velocity dispersions on the plane of the sky $\sigma_R/\sigma_T$) for the three projections (black, green and orange lines, for projection 1, 2 and 3, respectively). The shape of the velocity anisotropy profiles significantly depends on the particular choice of the line-of-sight.}
\label{fig:projections}
\end{figure*}
%---------------------------------------------------------------------------% 

So far we focused on characterizing the time evolution of the velocity anisotropy and its relation with the strength of the tidal field experienced by a cluster. In this section, we focus on the effect of the tidal field in the very outskirts of the clusters, specifically on the stars beyond the 99\% Lagrangian radii, that have been excluded in the above analysis. The periphery of a cluster is where we would expect the strongest impact of the tidal field, characterized by the presence of tidal tails, as well as the presence of potential stellar escapers, that are energetically unbound stars still residing within a cluster (\citealp{Kuepper2010,Claydon2017}).
\citet{Baumgardt2001} found that the fraction of potential escapers can be as high as 10-20\% of the cluster's bound stars, and becomes the dominant fraction of the cluster's stars beyond half of the Jacobi radius \citep{Kuepper2010}.
From an observational perspective, potential escapers are photometrically hard to disentangle from the cluster's real members at the same radial distance, and tidal tails can affect the properties of a cluster if they are looked at in projection. Note that all our simulations exhibit tidal tails at a 10 Gyr snapshot.

In Fig. \ref{fig:10kpc}, we compare the shape of the velocity anisotropy profiles of the accreted GCs at 10 kpc and the anisotropy profiles of the corresponding MW GC, for 10 Gyr snapshots. The velocity anisotropy profiles now include the outermost point corresponding to stars beyond the 99\% Lagrangian radius. The simulations show a sudden increase in $\beta$, from tangential anisotropy to radial anisotropy in the very outskirt, indicative of the presence of the tidal tails, for both the accreted and the MW GCs.\footnote{Due to the different cut-off prescription for the simulations of GCs solely in the MW (as noted in Section \ref{sec:3}), the increase in the outer parts for these simulations is weaker.}

Given the significance of the increase in anisotropy in the outskirts, we wish to test whether, from an observational perspective, this could influence the measured values of velocity anisotropy while looking at a cluster in projection. In Fig. \ref{fig:projections} we present three projected views of the 10 Gyr snapshot of the simulations DWL-MW10-evap, on the x-y (projection 1), x-z (projection 2) and z-y (projection 3) planes, with the x-y plane corresponding to the orbital plane of the cluster. For each projection we calculate the velocity anisotropy profile binned in projected Lagrangian radii (i.e. cylindrical radii along the line-of-sight containing a specific percentage of the total mass), both for the $\beta$ parameter and the ratio of the projected radial and tangential velocity dispersions on the plane of the sky $\sigma_R/\sigma_T$ (as usually measured with proper motion datasets). The projected Lagrangian radii considered are the one between 40\% and 60\%, 60 and 80\%, 80\% and 90\%, 90\% and 99\%, and beyond 99\% of the cluster's total mass. The inner points correspond to values of anisotropy within the 50\% projected Lagrangian radius. Note that these Lagrangian radii differ for each projection adopted.

Figure \ref{fig:projections} shows that the shapes of the projected velocity dispersion profiles depend on the chosen line-of-sight. In projection 1, the tidal tails are looked at from above and do not contaminate the line-of-sight, whereas in the other projections, they do provide a significant source of contamination. In particular, when measured with the $\beta$ parameter, projection 2 and 3 give rise to a more isotropic/more tangential velocity distribution, respectively, due to the particular orbital configuration of the tidal tails. Note that these differences are already significant for regions right beyond the projected half-mass radius. The contamination is instead limited to the outermost radii when the anisotropy is measured with $\sigma_R/\sigma_T$. This is particularly evident in projection 3, where a strong value of radial anisotropy emerges in the outer bin. Since the tidal tails in this projection are not directly visible (because they are looked at along the line-of-sight), we suggest that this kinematic feature in the anisotropy profile could be used as a possible signature of the presence of tidal tails when they are not photometrically detected. Figure \ref{fig:projections} suggests that the interpretation of present and upcoming data enabling the direct measurement of velocity anisotropy (e.g. \citealp{Bellini2014,Pancino2017}) should pay particular attention to the interplay between projection effects and possible presence of tidal tails along the line-of-sight.

%---------------------------------------------------------------------------------------------------------------------------%
%  			(5)	Discussion and Conclusions
%---------------------------------------------------------------------------------------------------------------------------%
\section{Discussion and Conclusions}
\label{sec:discussion}

In this paper we explored the possibility of detecting globular clusters that underwent an accretion origin based on their internal kinematic properties, namely the anisotropy in velocity space. Our simulated clusters all start with 3 Gyr of evolution on a circular orbit in a dwarf galaxy that is at a certain distance from the MW centre. After 3 Gyr the clusters experience a time-dependent tidal field that represent either the stripping of the dwarf galaxy while it orbits the MW potential, or the increase of the tidal field due to the infall of the dwarf-GC system. These two scenarios, Dwarf Falls and Dwarf Evaporates, allowed us to explore the main physical ingredients taking part in a dwarf galaxy-MW merger, with different initial configurations and different tidal history.  

Our simulations showed that, initially, the velocity anisotropy is determined by the tidal field of the host dwarf galaxy, and while the accretion process takes place (corresponding to an increasingly more dominant MW tidal field relative to the one of the dwarf galaxy), the clusters will start to adapt to the new tidal environment. At the end of the accretion phase, accreted clusters can either exhibit more radial or less radial anisotropy than MW globular cluster at the same distance, if they have experienced significantly weaker or significantly stronger tidal field until this time, respectively. These differences are stronger for the intermediate-outer radii of the clusters. However they do not represent distinctive and unique signatures since a variety of other effects could produce degenerate results, for example eccentric orbits around the MW or disc shocking. Moreover, at 10 Gyr, the differences in velocity anisotropy are quickly erased since, in a few relaxation times, the clusters fully respond to the new tidal environment.

When we focused on the very outskirts of the clusters (stars beyond the 99\% Lagrangian radius), we observed a sharp increase of the velocity anisotropy toward strong values of radial anisotropy. We showed that this effect is due to the presence of tidal tails forming in all our simulations at 10 Gyr. We further showed, from an observational perspective, that the presence of tidal tails can significantly contaminate the measurements of velocity anisotropy in the outer regions when the clusters are observed in projections. We suggested that the contamination of the velocity anisotropy profiles could be used as a signature of the presence of tidal tails.

Our kinematic study is consistent with the previous work of \citet{Miholics2016}, where a subset of the simulations used in this work were originally exploited to determine morphological differences between accreted and MW globular clusters. Also on the morphological point of view (specifically the size of the half-mass radius), clusters that underwent accretion will have their size determined by whichever tidal field is the strongest at any point in time. Once the MW tidal field become dominant, the cluster will adapt to the new potential and quickly becomes the same size of a cluster that has evolved solely in the MW (see also, \citealp{Miholics2014,Bianchini2015b}).

Beside the study of the differences and similarities of the velocity anisotropy that accreted and MW globular clusters acquire throughout their evolution, our analysis allowed us to characterize the details of the development of anisotropy under different time-dependent tidal environments. Specifically, at 10 Gyr, our simulations show that the velocity anisotropy profiles span a wide range of values: they are in general isotropic in the inner region and can develop, starting from $\simeq r_{50}$, either radial or tangential anisotropy that further increases in the outer parts, or remain isotropic throughout the radial extension. 
Independently of the accreted origin, the specific value of velocity anisotropy at 10 Gyr primarily depends on the strength of the tidal field cumulatively experienced by a cluster. Radial anisotropy is associated to weaker tidal fields, while more isotropic velocity distributions and tangential anisotropy are associated to progressively stronger tidal fields that are able to preferentially strip low-mass stars in radial orbits. This corresponds to clusters with increasingly more tidal filling configurations ($r_{50}/r_j\gtrsim0.17$), that have experienced higher mass loss ($\gtrsim60\%$) and with relaxation times $t_{rel}\lesssim10^9$ Gyr. 

Our characterization of the time evolution of the velocity anisotropy provides a theoretical benchmark for the understanding and interpretation of the large amount of three-dimensional internal kinematic data that are progressively becoming available for large samples of MW globular clusters (e.g., HST proper motions, e.g. \citealp{Bellini2014}, \citealp{Watkins2015}, and Gaia data, e.g. \citealp{Pancino2017}) that make it possible to directly measure the velocity anisotropy for both the inner and outer regions of Galactic globular clusters. Any significant deviations from our predictions could be used to point out more complex dynamical histories (e.g. tidal shocking, stripped nuclei of a dwarf galaxy), primordial fingerprints of formation that have not been considered in the initial conditions of our simulations (e.g. primordial violent relaxation), or effects due to non-standard mass functions.

\section*{Acknowledgments}
We thank Anna Lisa Varri and the referee for useful comments. PB acknowledges financial support from a CITA National Fellowship. 

\bibliographystyle{mnras} % style aa.bst
\bibliography{biblio} % your references Yourfile.bib

\end{document}